\documentclass{PoS}

\title{Search for the Higgs boson in the $\gamma \gamma$ final state at the Tevatron}

\ShortTitle{$H\to\gamma \gamma$ at the Tevatron}

\author{\speaker{Krisztian Peters}\thanks{for the CDF and D0 Collaborations}\\
        School of Physics and Astronomy, The University of Manchester,\\
Oxford Road, Manchester M13 9PL, United Kingdom\\
        E-mail: \email{petersk@fnal.gov}}

\abstract{We present searches for Higgs bosons decaying to the di-photon final state 
using up to 5.4 fb$^{-1}$ of data at a center-of-mass energy of $\sqrt s =1.96$ TeV at the 
Fermilab Tevatron collider. Whilst the branching ratio to the di-photon final 
state is small in the Standard Model, this channel contributes appreciably 
to the overall Higgs sensitivity at the Tevatron. In parallel, the limit is re-interpreted 
in fermiophobic models where the di-photon branching ratio is considerably larger. 
This decay channel will be of major importance in the light mass Standard Model 
Higgs search at the LHC.}

\FullConference{35th International Conference of High Energy Physics - ICHEP2010,\\
		July 22-28, 2010\\
		Paris France}

\begin{document}

\section{Introduction}

The existence of the Higgs boson is the only fundamental particle of the 
Standard Model (SM) which has still to be confirmed. Its observation would be 
a crucial step in establishing the mechanism of electroweak symmetry breaking
and mass generation. A lower limit of 114.4 GeV was set by 
the LEP experiments on the mass of the Higgs boson \cite{lep}, while 
the Tevatron experiments recently excluded the mass range between 158 and 
175 GeV \cite{TeV}. In addition, an indirect upper limit of 158 GeV can be inferred 
from precision electroweak data \cite{ew}. 
These limits and those given below are all defined at the 95\% C.L.

These results indicate that if the SM is the correct underlying theory, 
a light Higgs boson should exist with a mass within reach of the Tevatron collider, 
which is below $\sim$190 GeV.  The overall search strategy at the Tevatron is to 
investigate different production mechanisms and a large number of final 
states to scan this whole mass range. At the lower Higgs mass range, below $\sim$135 GeV, 
the main search channels are the associated productions with vector bosons where the Higgs 
decays to $b\bar b$ pairs. For masses above $\sim$135 GeV the search is mainly focused on the gluon fusion 
production process where the Higgs decays to $WW$ pairs.

The sensitivities of these searches are enhanced by additional channels, among them is the Higgs decay 
to di-photons.  Within the SM the BR($H\to\gamma\gamma$) is suppressed by more than two orders of magnitude
 compared to the BR($H\to b\bar b$), 
which results in a very small di-photon production rate. Despite this, $H\to\gamma\gamma$ 
contributes appreciably to the overall 
Higgs sensitivity as it has several advantages over the $b\bar b$ final states. 
Since the di-photon selection reduces the (otherwise overwhelming) 
multijet background significantly, the gluon fusion production mode can be exploited.
This gives already an order of magnitude enhancement compared to the 
Higgs produced in association with a vector boson.  
Further advantage can be taken from the precise photon energy measurements. 
The di-photon invariant mass resolution is below 3 GeV, more than five times better 
than the dijet mass resolution of the $b\bar b$ decay at the Tevatron experiments. Due to these advantages, $H\to
\gamma\gamma$ provides important 
additional sensitivity to the overall Tevatron result especially in the difficult intermediate mass 
region of $\sim$130 GeV where the 
BR($H\to\gamma\gamma$) has its largest contribution. Searches for $H\to\gamma\gamma$ 
at the Tevatron are also forerunners to similar planned searches at the LHC. 

For certain beyond the SM scenarios significant 
enhancements to the production rate would  be possible. This can be realized with new 
particles that might affect the loop mediated $gg\to H$ and H$\to\gamma\gamma$ couplings. 
Furthermore, the BR($H\to\gamma\gamma$) can be increased in models with modified Higgs 
couplings to fermions. Such models were discussed in the literature \cite{fermio} and they can 
be probed with H$\to\gamma\gamma$ searches at the Tevatron. In this note, we will discuss the 
so-called fermiophobic Higgs scenario, where the Higgs couplings to all fermions are suppressed. 
In this model, around Higgs masses of 110~GeV, the BR($H\to\gamma\gamma$) is by almost two 
orders of magnitude enhanced compared to the SM value.

\section{Search strategy}

The general strategy at the Tevatron experiments is to perform the H$\to\gamma\gamma$ search
as model independent as possible. This is achieved with an inclusive selection, 
where only the di-photon invariant mass observable is used
to look for a bump in the deeply falling spectrum. This way the signal acceptance and analysis sensitivity is 
basically independent of the Higgs production mechanism. In general, such a search can probe 
for any narrow resonance decaying into di-photons in a quasi model-independent way. 
In the case of the SM Higgs searches, the main production 
mechanism is gluon-fusion production, however the associated production with vector bosons and 
the vector boson fusion (VBF) production mechanism provide $\sim$30\% additional signal to the SM searches 
and are also considered.

As the event signature of this search is simply two photons in the calorimeter, the relevant aspects for this search are 
a good calorimeter resolution, efficient photon identification algorithms and well understood background models 
(which are obtained with data driven techniques).

Both the CDF and D0 calorimeters have a central ($|\eta | < 1.1$) and a forward section. The CDF calorimeter 
uses scintillating tiles with lead as absorber material in the electromagnetic (EM) section and is nearly free of noise. D0  
has a Liquid Argon calorimeter with mostly uranium as absorber and is finely segmented with $\sim$60k cells in total. 
Both calorimeters are calibrated regularly with special triggered data and have comparable EM resolutions with 
$\sim$2\% constant term. 

 After the cell level calibration, detailed object level calibrations are performed to take into account specifics of 
 photons. As an example, the amount of material in front of the D0 calorimeter has significantly increased
 due to the addition of the silicon vertex detector and the solenoidal magnet during the upgrade. In consequence, the shower 
 maximum of EM particles moved towards the frontal calorimeter layers, the EM response and resolution depend 
 more strongly
 on the particle energy and incident angle and different energy-loss corrections have to be applied to electrons and 
 photons. The D0 analysis could benefit from the detailed understanding that was achieved and on the tools that were 
 developed for the $W$ mass measurement \cite{Wmass}. Energy-loss corrections were measured in $Z\to ee$ events 
 and propagated to different energy scales and to photons with a tuned GEANT simulation.

Both experiments select photons from EM clusters with the following criteria:
(1) High EM fraction or cluster located in shower maximum detector, (2) Isolated in the calorimeter, (3) Isolated in the tracker,
(4) Transverse shower profile consistent with EM objects, (5) No associated tracks and no 
pattern of hits consistent with electrons. Differences between data and simulation are calibrated using photons 
from radiative $Z$ decays ($Z\to\ell\ell\gamma$) and $Z\to ee$ events. At D0 the photon purity is further improved 
with a five-variable Artificial Neural Network (NN) which uses mainly isolation related variables 
from the tracker, pre-shower and calorimeter. This NN is trained using QCD $\gamma\gamma$ and 
di-jet Monte Carlo samples (MC). 
Its performance is verified with $Z\to\ell\ell\gamma$ data events which shows an excellent agreement between data and MC.
The H$\to\gamma\gamma$ search requires NN$>0.1$. This cut is almost 100\% efficient for photons while rejecting 
50\% of misidentified jets.

\section{Search for the SM Higgs}

The data for di-photon searches is collected with a suite of calorimeter only triggers. These are di-EM 
triggers with a $p_{T}$ threshold that varies between 12 and 25 GeV.  The CDF 
analysis uses in addition two single-photon triggers with a threshold of 50 and 70 GeV. For both experiments the 
trigger efficiency is very close to 100\% after the offline selection cuts. A primary vertex is identified 
within the acceptance of the tracking detectors; the photon kinematics is then computed with respect to this vertex. 
Two photon candidates are required in the central calorimeters, which satisfy the photon selection 
criteria as described in the previous section, 
with $p_{T}$ above 15 (25) GeV and a di-photon invariant mass above 30 (60) GeV for CDF (D0). 

All backgrounds have steeply falling di-photon mass spectra with different slopes. The main reducible 
backgrounds are either electrons misidentified as photons from Drell-Yan production (DY) or jets misidentified as 
photons from QCD di-jet or photon+jet production. In the D0 analysis the DY part is estimated with MC 
normalized to the NNLO 
theoretical cross section \cite{DY}. QCD di-jet and photon+jet production are directly estimated from data by 
using a 4 x 4 matrix background estimation method \cite{D0publ}. The main irreducible background is direct QCD 
di-photon production which is challenging to predict theoretically. This background is estimated from 
fit to data in the invariant mass distribution after subtraction of the reducible backgrounds. The fit 
range is [70, 200] GeV, excluding the 
signal region, which is defined to be the interval of the Higgs mass $m_{H} \pm 15$ GeV. In the CDF analysis 
the sum of all backgrounds together is taken from an inclusive sideband fit method (similar to the di-photon estimate 
in the D0 analysis).  

Systematic uncertainties affecting the normalization and shape of the di-photon spectrum are 
estimated for both signal and backgrounds. The main uncertainties are (D0 example): 
the integrated luminosity (6\%), acceptance 
due to the photon identification efficiency (7\%), electron misidentification rate (19\%),  DY cross section 
(4\%) and PDFs for signal (2\%). Systematic uncertainties have a small effect on limits and the 
final sensitivity is still driven by the available statistics.

\begin{figure}[t]
\centering
\includegraphics[width=70mm]{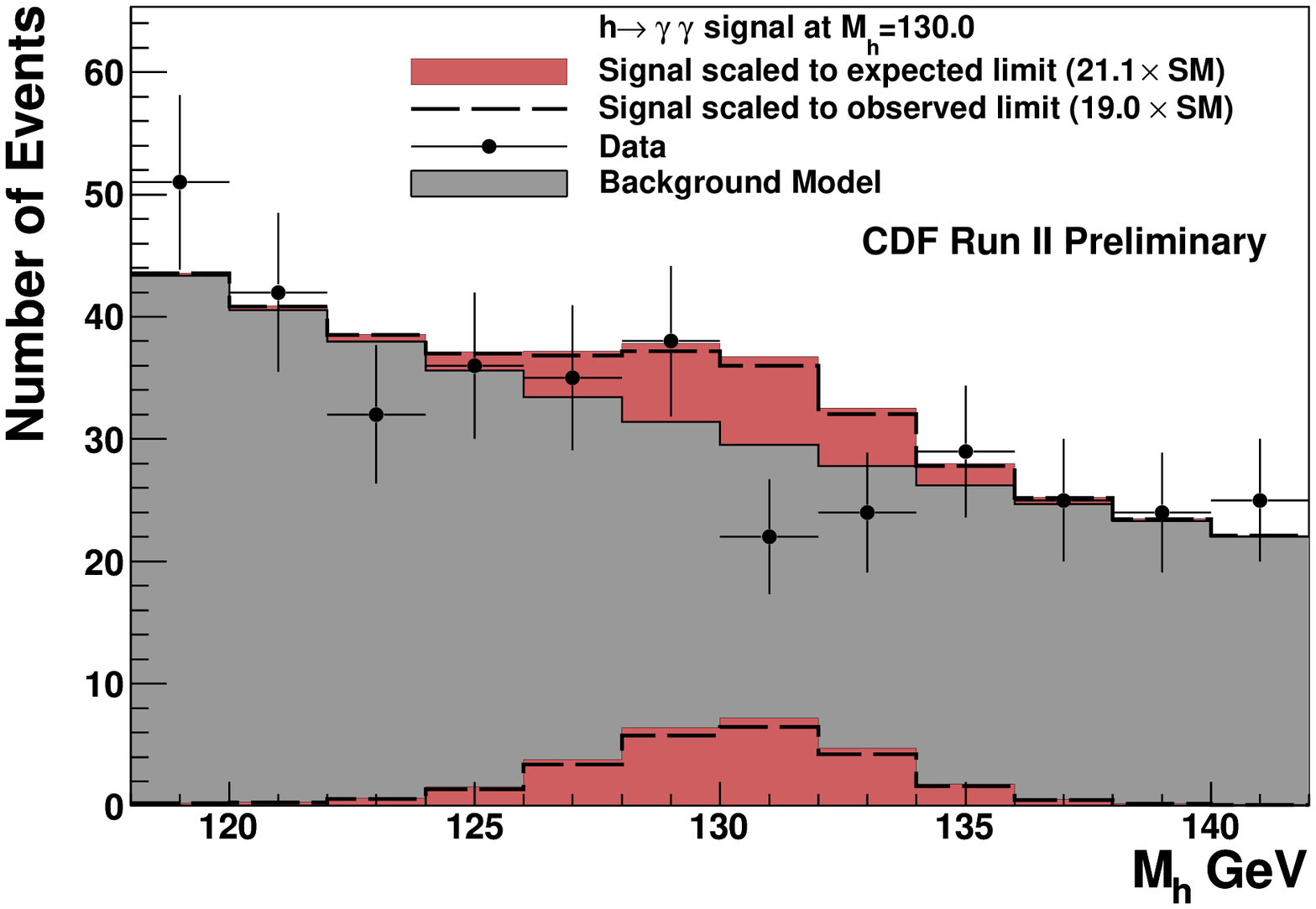}
\includegraphics[width=76mm]{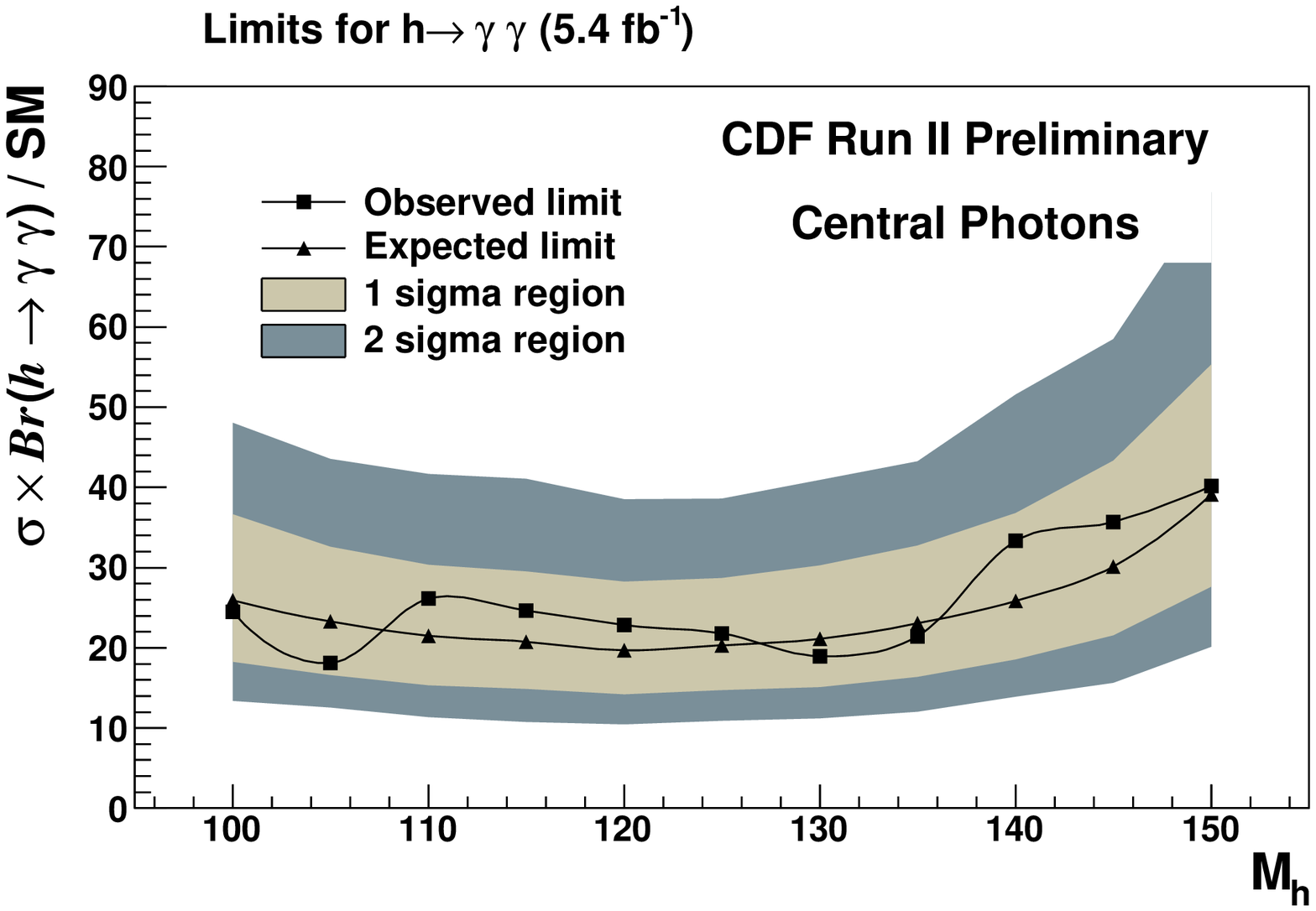}
\caption{$M_{\gamma\gamma}$ spectrum in the search region for $m_{H} = 130$ GeV and upper limits on the production cross section times branching ratio relative to the Standard Model prediction in the CDF analysis.}
\end{figure}

Fig.~1 illustrates the $M_{\gamma\gamma}$ spectrum in the search region for $m_{H} = 130$ GeV (CDF example), 
which is in 
good agreement with the background prediction. The $M_{\gamma\gamma}$ spectrum in the search region 
is used to derive upper limits on the production cross section times branching ratio for $H\to\gamma\gamma$ as a 
function of the Higgs mass. With the current datasets the limits are, for both the CDF and D0 analyses, a 
factor of $\sim$20 above the 
SM expectation for the Higgs mass range between 100 and 140 GeV \cite{CDFgaga, D0gaga}. 

These results add a 5\% sensitivity to the overal SM Higgs combination at the Tevatron and are especially important
for the mass region around 130 GeV. It is expected that further significant improvements can be 
achieved in the future by moving from the inclusive selection to a dedicated optimized search using multivariate techniques.  

\section{Search for the fermiophobic Higgs}

In the fermiophobic scenario the BR($H\to\gamma\gamma$) would be largely enhanced as discussed in Section~1. Since 
in this scenario the Higgs does not couple to fermions the gluon-fusion production mode is absent. In the associated 
production with vector bosons and in the VBF production Higgs bosons decaying to photons are produced with 
a significant recoil. This is exploited by requiring a significant di-photon transverse momentum. The rest of the 
event selection and the analysis procedure are very similar to the SM production searches. 

Fig.~2 shows limits on the BR($H\to\gamma\gamma$) as a function of fermiophobic Higgs mass (D0 example).  
Within the fermiophobic scenario, Higgs masses of 106 (102.5) GeV are excluded at CDF (D0) \cite{CDFgagaF, D0gagaF}. 
These results are close to the mass exclusion of 109.7 GeV from the combined four LEP experiments, and a 
Tevatron combination would potentially exceed this limit.  Furthermore, limits on BR($H\to\gamma\gamma$) at the Tevatron are probing 
new territory beyond the kinematic reach of LEP above $\sim$125 GeV.

\begin{figure}[t]
\centering
\includegraphics[width=80mm]{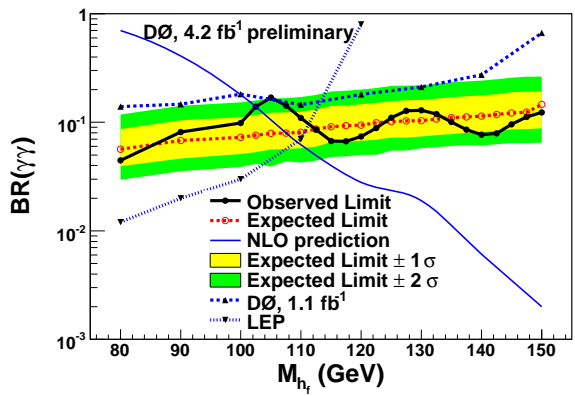}
\caption{Limits on BR($H\to\gamma\gamma$) as a function of fermiophobic Higgs mass in the D0 analysis.}
\end{figure}

\end{document}